\documentclass[journal=jacsat,manuscript=article,layout=twocolumn]{achemso}

\usepackage{amssymb}
\usepackage{graphicx}
\usepackage{dcolumn}
\usepackage[version=3]{mhchem}
\usepackage{textcomp, gensymb}
\usepackage{chemformula}
\usepackage[fontsize=10pt]{fontsize}

\DeclareUnicodeCharacter{2212}{$-$}

\usepackage{xr}
\makeatletter
\newcommand*{\addFileDependency}[1]{
  \typeout{(#1)}
  \@addtofilelist{#1}
  \IfFileExists{#1}{}{\typeout{No file #1.}}
}
\makeatother

\newcommand*{\myexternaldocument}[2][]{
  \externaldocument[#1]{#2}
  \addFileDependency{#2.tex}
  \addFileDependency{#2.aux}
}

\myexternaldocument{SI}

\author{Vivian R. K. Wall}
\affiliation{Department of Chemistry, University of California, Berkeley, CA 94720, USA}

\author{Andrew Yang}
\affiliation{Department of Applied Physics and Applied Mathematics, Columbia University, New York, NY 10027, USA}
\alsoaffiliation{Present Address: Department of Applied Physics and Materials Science, California Institute of Technology, Pasadena, CA 91125, USA}

\author{Bryan A. Sanchez Monserrate}
\affiliation{Department of Chemistry,Stony Brook University, Stony Brook, NY 11794, USA}

\author{Ahhyun Jeong}
\affiliation{Department of Chemistry, University of Chicago, Chicago, IL 60637, USA}

\author{Christian P. N. Tanner}
\affiliation{Department of Chemistry, University of California, Berkeley, CA 94720, USA}

\author{Joshua Portner}
\affiliation{Department of Chemistry, University of Chicago, Chicago, IL 60637, USA}

\author{John P. Davis}
\affiliation{Department of Chemistry, University of California, Berkeley, CA 94720, USA}

\author{Tim B. van Driel}
\affiliation{Linac Coherent Light Source, SLAC National Accelerator Laboratory, Menlo Park, CA 94025, USA}

\author{Burak Guzelturk}
\affiliation{Advanced Photon Source, Argonne National Laboratory, Lemont, IL 60439, USA}

\author{Gihan Kwon}
\affiliation{National Synchrotron Light Source II, Brookhaven National Laboratory, Upton, NY 11973, USA}

\author{Olaf J. Borkiewicz}
\affiliation{Advanced Photon Source, Argonne National Laboratory, Lemont, IL 60439, USA}

\author{David T. Limmer}
\affiliation{Department of Chemistry, University of California, Berkeley, CA 94720, USA}

\author{Dmitri V. Talapin}
\affiliation{Department of Chemistry, James Franck Institute, and Pritzker School of Molecular Engineering, University of Chicago, Chicago, IL 60637, USA}
\alsoaffiliation{Center for Nanoscale Materials, Argonne National Laboratory, Lemont, Illinois 60439, United States}

\author{Karena W. Chapman}
\affiliation{Department of Chemistry, Stony Brook University, Stony Brook, NY 11794, USA}

\author{Simon J. L. Billinge}
\affiliation{Materials Department, University of California, Santa Barbara, CA 93106, USA}
\alsoaffiliation{Department of Applied Physics and Applied Mathematics, Columbia University, New York, NY 10027, USA}

\author{Naomi S. Ginsberg}
\email{nsginsberg@berkeley.edu}
\affiliation{Department of Chemistry, University of California, Berkeley, CA 94720, USA}
\alsoaffiliation{Department of Physics, University of California, Berkeley, CA 94720, USA}
\alsoaffiliation{Molecular Biophysics and Integrated Bioimaging Division, Lawrence Berkeley National Laboratory, Berkeley, CA 94720, USA}
\alsoaffiliation{Materials Sciences and Chemical Sciences Division, Lawrence Berkeley National Laboratory, Berkeley, CA 94720, USA}
\alsoaffiliation{Kavli Energy NanoSciences Institute, University of California, Berkeley, CA 94720, USA}
\alsoaffiliation{STROBE, NSF Science \& Technology Center, Berkeley, CA 94720, USA}

\title{Pair distribution function oscillations at electrostatically stabilized nanocrystal surfaces reflect nanocrystal lattice periodicity}

\begin{document}

\begin{abstract}
Nanocrystals (NCs) stabilized electrostatically by short, multivalent inorganic ligands exhibit rich interactions and phase behavior believed to hinge on the configuration of ions at the NC surface, yet these molecular details have not previously been measured. Here we use double-difference pair distribution functions (ddPDFs) of total X-ray scattering to determine the configuration of \ch{Sn2S6^4-} ligands on PbS NCs suspended in multivalent electrolyte. The measured ddPDFs exhibit  a $\sim$3-\AA\ wavelength oscillation that extends  $\sim 10$~\AA. Comparison with a  PbS NC model decorated with \ch{Sn2S6^4-} ligands indicates ligands are aligned $\sim$radially, suggesting bidentate binding. Calculated partial PDFs show that the oscillatory signal is dominated by nanocrystal-Pb to ligand-Sn correlations. A simplified model that replaces the ligand with a strongly scattering Pt atom near the NC surface confirms  the oscillation period in the ddPDF originates from the NC lattice constant. Nanocrystal lattice periodicity therefore merits consideration when interpreting ddPDFs of colloidal NCs.

\end{abstract}

\subsubsection{Keywords:}
nanocrystals, 
electrostatic stabilization,
X-ray scattering, 
pair distribution function,
solvation shell
\subsubsection{}
\maketitle
%
Nanocrystals (NCs) that are colloidally stabilized electrostatically by short, multivalent inorganic ligands on their surface, rather than typical, bulkier organic ones 
are a promising class of nanoscale materials.\cite{kovalenko_colloidal_2009, kovalenko_expanding_2010,protesescu_atomistic_2015, lee_band-like_2011,tanner_situ_2024,hurley_situ_2024,tanner_origins_2025,jeong_colloidal_2024} 
These NCs are suspended in polar solvents, and their interactions can be controlled by introducing excess salt composed of the same multivalent ions that are bound to 
the NC surface. 
Such systems exhibit rich phase behavior: 
for example, they can undergo a nonclassical, two-step crystal formation process through a metastable liquid intermediate,
\cite{coropceanu_self-assembly_2022,tanner_enhancing_2025} wherein the NCs exhibit suppressed self-diffusion.\cite{tanner_origins_2025} 
These behaviors are believed to hinge on the configuration of the bound and unbound multivalent ions around the NCs. 
Simulations have shown that a high ratio of NC dielectric constant to solvent dielectric constant generates layering in the NC solvation shell.\cite{jeong_atomic_2026,guerrero_garcia_polarization_2014} 
Despite the crucial role that the environment around electrostatically-stabilized NCs plays in dictating their interactions, these molecular details have not previously been studied experimentally. 
It remains to be determined how the multivalent ligands are arranged on the NC surface, and whether ions beyond the bound ligands are organized into the predicted layers.

\begin{figure}[tbh]
    \centering
    \includegraphics[width=9cm]{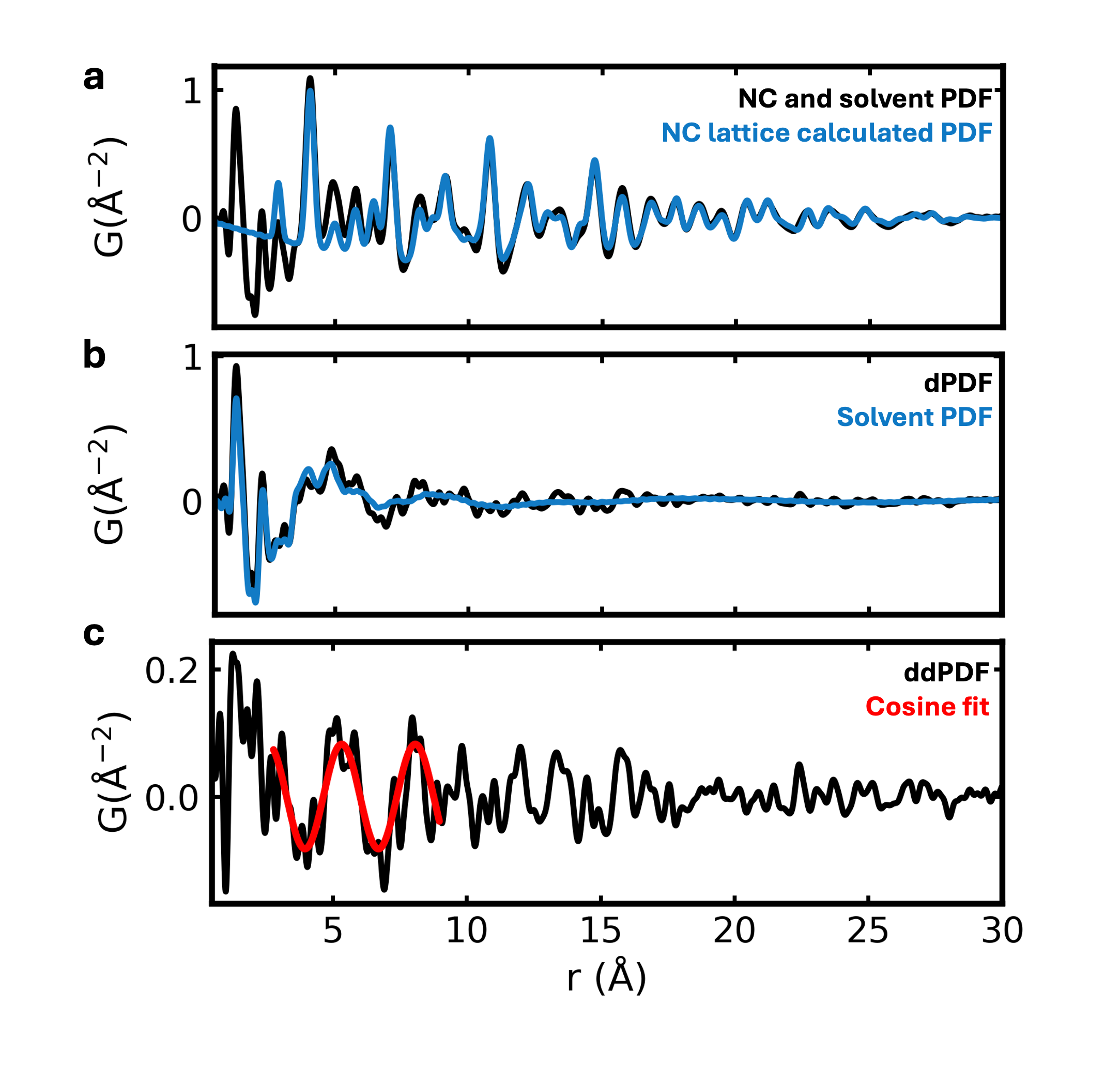}
    \caption[ddPDF calculation.]{Double-difference PDF calculation. \textbf{a.} PDF of PbS NCs in 60:40 NMF:DMF with no added K$_4$Sn$_2$S$_6$ (black) and fitted PDF of PbS NC lattice model (blue). \textbf{b.} dPDF of PbS NC solution PDF minus PbS NC model PDF (black) and PDF of 60:40 NMF:DMF with no added K$_4$Sn$_2$S$_6$ matched to dPDF using diffpy.morph (blue). \textbf{c.} ddPDF of dPDF minus morphed solvent PDF. The ddPDF is fit to a cosine function, yielding a period of $\sim$ 3 $\mathrm{\AA}$.
}
    \label{fig:differences}
\end{figure}

Measuring the molecular-scale structure at the surface of NCs is challenging. This molecular structure lacks long range order and has weak signal strength relative to those of the NC core and the bulk solvent. Pair distribution function (PDF) analysis of total X-ray scattering is a useful tool for studying structure without long range order.\cite{egami_underneath_2012} 
It has not, however, been applied to study the all-inorganic electrostatically-stabilized NCs in electrolyte solutions and the relationship to their phase behavior. 
Obtaining a reliable double difference PDF (ddPDF) is itself demanding: it is difficult to measure with adequate signal strength, signal-to-noise ratio, and reproducibility in the measurement to verify that the deduced PDF features are not misattributed. For example, previous ddPDF studies have been called into question or reevaluated to account for trace impurities' contributions to the signal,\cite{thoma_atomic_2019,thoma_ethanolwater_2023} and differencing schemes are 
susceptible to over- or under-subtraction and suffer from an amplified noise signal that can be misinterpreted. 
Interpretation presents a further challenge. Oscillatory features in ddPDFs have generally been attributed to the size and packing of the molecules in the shell,\cite{zobel_universal_2015,kamysbayev_nanocrystals_2019} while the possible contribution of the periodic lattice of the NC core itself has not yet been considered.

In this work we use ddPDFs to uncover the molecular-scale configuration of multivalent ions around electrostatically-stabilized PbS NCs in electrolyte solution. 
The large scattering cross-sections of Pb and Sn raise the surface-related signal above that of the residuals that limit the subtractions. Furthermore, the rigidity and bidentate binding of the \ch{Sn2S6^4-} ion narrow the distributions of bound-ligand distances and orientations relative to those of more conformationally flexible organic ligands and the unbound molecules beyond the ligand layer.\cite{protesescu_atomistic_2015} 
Our ddPDFs show $\sim3$~{\AA} oscillations that persist to $\sim 10~${\AA} (\textbf{Figure \ref{fig:differences}}). We model this system with a truncated octahedral PbS NC, decorated with \ch{Sn2S6^4-} ligands on the (111) facets. The calculated differential PDF of this simple model matches well with the experimental ddPDF. By identifying the ligand positions and orientations in calculated PDFs that best match the experimental ddPDFs, we infer that ligands are arranged roughly perpendicular to the NC surface. 
We further find that a single layer of ligands accounts for the signals in the measured ddPDF, while the ddPDF is only weakly sensitive to molecules beyond this layer, which suggests that they are already significantly disordered. 
Calculated partial PDFs show that the strongly scattering Pb-Sn correlations dominate the ddPDF, so that the measurement principally reports the positions of the ligand Sn atoms relative to the underlying periodic Pb sublattice. This interpretation is supported by a simple model of a single Pt atom at the surface of the PbS NC that 
shows that the onset of the oscillations reflects the distance of the atom from the NC surface while their period reflects the NC lattice constant. 
These features of the electrostatically-stabilized NC system facilitate the attribution of the ddPDF oscillations to the NC lattice in this study, yet we suggest that lattice periodicity merits investigation alongside molecular size and packing when interpreting oscillatory ddPDFs of other colloidal NCs as well.

PDFs of electrostatically-stabilized NC solutions were extracted from total X-ray scattering measured at high energies. We measured total X-ray scattering of solutions of 5.9$\pm$0.4 nm PbS NCs (\textbf{Figure \ref{fig:PDFSAXS}a}) with \ch{Sn2S6^4-} ligands in 60:40 v/v mixtures of N-methylformamide (NMF) to N,N-dimethylformamide (DMF).\cite{coropceanu_self-assembly_2022} In some cases we added a small amount of additional K$_4$Sn$_2$S$_6$ to the solutions. These concentrations ($<30$ mM) were not enough to quench the NCs to a condensed phase.\cite{tanner_enhancing_2025} We also measured total X-ray scattering of the pure solvent mixture and the solvent with added K$_4$Sn$_2$S$_6$. Similarly, total scattering of solutions of CdSe NCs with \ch{Sn2S6^4-} ligands and relevant solvent and salt backgrounds were also measured. The Fourier transform of the reduced structure factor of the total scattering yields a pair distribution function, $G(r)$ (\textbf{Figure \ref{fig:IFG}}). To obtain the ddPDF, first a calculated $G(r)$ for the NC lattice correlations is subtracted from the $G(r)$ of the NC solution (\textbf{Figure \ref{fig:differences}a}). Subsequently, the appropriate solvent background $G(r)$ is subtracted  (\textbf{Figure \ref{fig:differences}b}), leaving only correlations that change or appear in the solution such as the correlations associated with the NC-solvent interface, including the \ch{Sn2S6^4-} ligands  (\textbf{Figure \ref{fig:differences}c}). These subtractions were performed with the help of the \textsc{diffpy.morph} package\cite{yang_diffpymorph_2026}. For sample, measurement, and analysis details, see the supporting information.

The PbS ddPDFs exhibit oscillations on an intermediate length scale of $\sim$2-10 $\mathrm{\AA}$. 
A fit to a cosine function, $G(r)=A_{\mathrm{c}}$cos$(k_{\mathrm{c}}r+\phi_{\mathrm{c}})$ is shown in \textbf{Figure \ref{fig:differences}c}, yielding a period of $\sim$3 $\mathrm{\AA}$. 
The oscillations maintain their amplitude throughout this range. Beyond $\sim$10 $\mathrm{\AA}$, there are small peaks introduced from the finite $Q$-range 
of the Fourier transform or small residuals from the background subtraction. This ddPDF profile was reproduced from total scattering collected in a separate experiment with a different batch of NCs and different measurement conditions  (\textbf{Figure \ref{fig:variations}a}).

To further analyze the ddPDF we developed a model of a truncated octahedral PbS NC decorated with \ch{Sn2S6^4-} ligand ions from which we calculated the PDF. Before building this full model, we first more 
efficiently identif
ied 
plausible ligand configurations 
by screening a
preliminary model consisting of a single \ch{Sn2S6^4-} ion near a small slab of PbS. 
This involved randomly sampling the ion's orientation and distance relative to the slab and scoring each calculated PDF against 
the experimental ddPDF 
over the range of 3-8~{\AA}, where the most prominent oscillatory features in the experimental ddPDF are observed (\textbf{Figure \ref{fig:model}}). 
The single-ion PDF 
shares its general profile with the experimental ddPDF 
but fails to capture its broad features in 
this range, suggesting that 
the disorder arising from 
a finite number of ligands distributed over the NC surface must also be captured. 
We therefore retained the contiguous regions of configuration space with mean Pearson correlation coefficient (PCC)\cite{witte_statistics_2017} above the 85th, 90th, and 95th percentile (\textbf{Figure \ref{fig:model}a}) as the basis for the full NC model. The PCC is calculated according to \textbf{Equation \ref{eq:PCC}} in the Supporting Information, where PCC = 1 is fully correlated and PCC = 0 is uncorrelated.   

To create this full NC model, a PbS truncated octahedron NC\cite{hjorth_larsen_atomic_2017} with an effective diameter of $\sim$5.9 nm (\textbf{Figure \ref{fig:PDFSAXS}b}) is constructed with ligands on the surface. The appropriate number of ligand ions included is calculated to be $\sim$80 based on the known surface density of 0.75/nm$^2$\cite{portner_electrostatic_2023} and the surface area of the model NC. These 80 ions are divided evenly among the eight (111) faces, which have the highest density of Pb atoms and which are known to preferentially bind the \ch{Sn2S6^4-} ions, so that each (111) face binds 10~ligands. Neglecting the few ions that could be bound to (100) facets shows little impact (\textbf{Figure \ref{fig:100}}). The differential PDFs for these models are calculated by generating the partial PDFs for each element pair and summing all partial PDFs except those associated with the correlations of the atoms within the NC core. The calculated differential PDFs leave only the correlations at the surface of the NC, just as is the case in the experimental ddPDF. 

We generated three versions of the model for comparison (\textbf{Figure \ref{fig:percentiles}}), corresponding to using the 85th, 90th, and 95th percentile regions of the configuration space shown in \textbf{Figure \ref{fig:model}a}. The 90th percentile model strikes a balance, yielding the highest PCC relative to the experimental ddPDF in the range of 3-8 $\mathrm{\AA}$, PCC = 0.66. Because this model provides the best match, it is used to interpret the experimental data. \textbf{Figure \ref{fig:model}d} shows an illustration of the model\cite{momma_vesta_2011} and \textbf{Figure \ref{fig:model}e} shows the calculated differential PDF compared to the experimental ddPDF. The PDF is not meaningfully different when K$^+$ counterions are included (\textbf{Figure \ref{fig:K}}). 

\begin{figure*}[h]
    \centering
    \includegraphics[width=16cm]{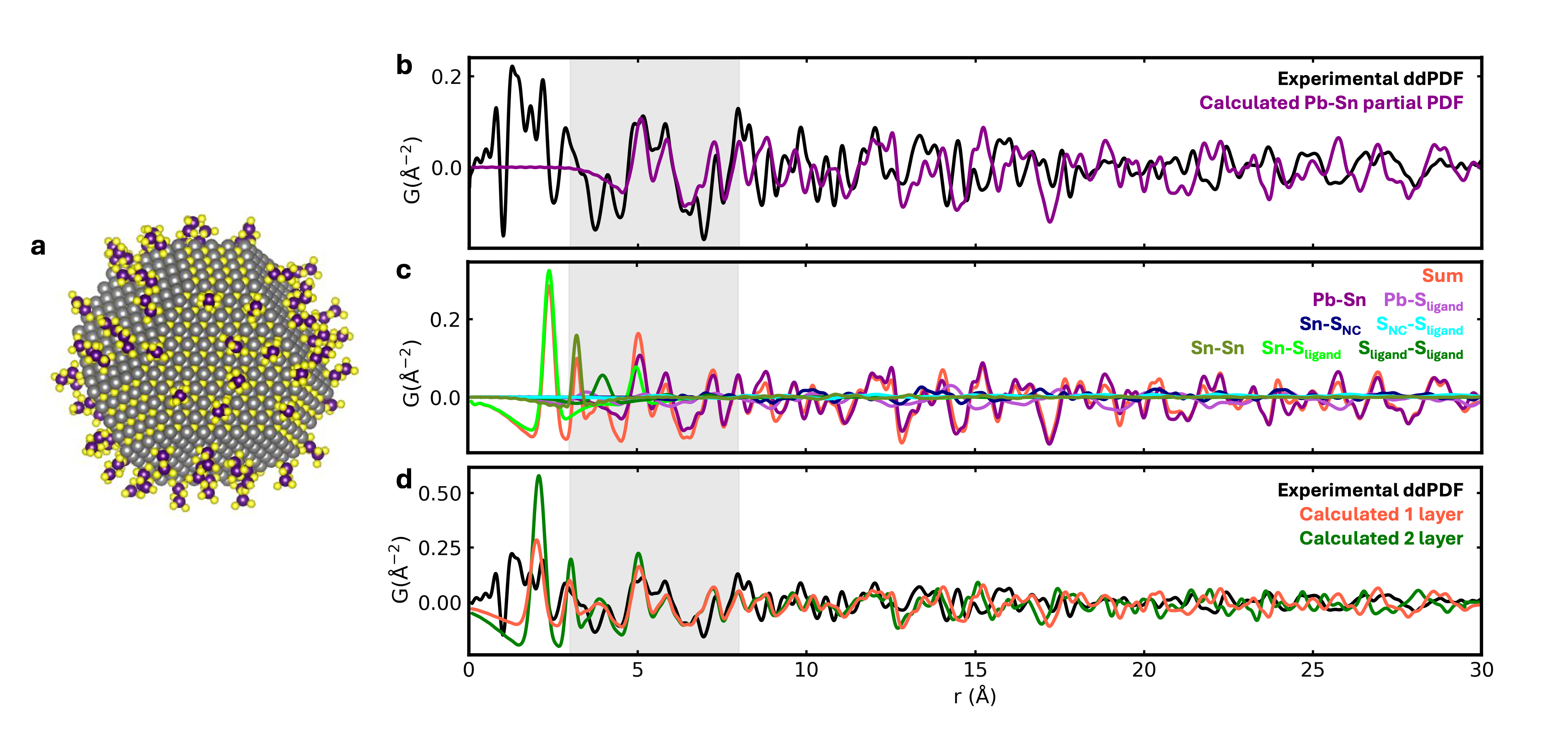}
    \caption[Pb-Sn correlations dominate the ddPDFs.]{Pb-Sn correlations dominate the ddPDFs. \textbf{a} Illustration of modeled NC with 80 ligands.\cite{momma_vesta_2011} \textbf{b.} Comparison of the experimental ddPDF with the calculated Pb-Sn partial PDF. The PCC between the two PDFs within the greyed region is 0.75. \textbf{c.} All NC-ligand and ligand-ligand calculated partial PDFs. The Pb-Sn partial PDF dominates the summed signal. \textbf{d.} Experimental ddPDF compared to calculated differential PDFs for model of NC with one and two layers of ions. The second layer was placed just beyond the first layer using the same tilt range as was used to construct the first layer. The calculated PDFs are rescaled to match the experimental ddPDF but their scaling relative to one another is not altered.
}
    \label{fig:modelvars}
\end{figure*}

The calculated PDF from the NC model with ligand ions matches the experimental ddPDF reasonably well (\textbf{Figure \ref{fig:model}e}).
Furthermore, this model is consistent with expectations, as evidenced by the strong match of the Pb-Sn correlations in the model to the measured ddPDF oscillations shown in \textbf{Figure \ref{fig:modelvars}a}. This match provides insight into the position and orientation of the ligands. The best distance between the Pb (111) surface of the NC and the terminal S atoms on the ligand ion is found to range from $\sim$3-3.5 $\mathrm{\AA}$, which is slightly longer than the estimated expected bond length between Pb and S.\cite{zeng_density_2005,doak_coherent_2012} The tilt of the ligand's long axis relative to the (111) surface normal is $\lesssim$ 25$^\circ$ and trends higher for smaller distances between the NC surface and the ligand center. This finding supports a \ch{Sn2S6^4-} ligand configuration of bidentate bonding to the Pb surface and roughly radial alignment.

Calculated partial PDFs reveal that the ddPDF signal is dominated by strongly scattering Pb-Sn correlations. The PCC between the experimental ddPDF and the calculated partial PDF for Pb-Sn in the range of 3-8 $\mathrm{\AA}$ is 0.75, which is even higher than that of the full calculated differential PDF. A comparison of the experimental ddPDF with the Pb-Sn partial PDF is provided in \textbf{Figure \ref{fig:modelvars}b}. \textbf{Figure \ref{fig:modelvars}c} shows the partial PDFs for all NC-ligand and ligand-ligand pairs. Each \ch{Sn2S6^4-} ligand in the model is identical, whereas a real system would contain some bond distortion and disorder. Thus, peaks associated with intramolecular bonds in the model ligand-ligand partial PDFs, like the Sn-S$_{\mathrm{ligand}}$ pair at $\sim$5 $\mathrm{\AA}$, are sharper than in the experimental case. Despite the inclusion of small atomic displacements in the model, the variation in intramolecular bond lengths is likely underestimated, and thus these peaks are artificially prominent. Even so, the Pb-Sn partial PDF dominates the summed signal, relative to the other partial PDFs. 

A single layer of ions in the model sufficiently captures the features in the experimental ddPDF (\textbf{Figure \ref{fig:modelvars}d}). Adding a second layer of ions to the model just beyond the first layer does not appreciably change the shape of the calculated differential PDF within the \textit{r}-range of the salient features in the experimental ddPDF. The orientations in the second layer are drawn from the same tilt range as the first layer (\textbf{Figure \ref{fig:model}a}).  Even though this strategy likely underestimates the tilt disorder in the second layer, it still shows minimal difference from the monolayer trace in the ddPDF. The PCC between the experimental ddPDF and the two layer calculated PDF in the range of 3-8~{\AA} is 0.61, which is similar to that of the single layer. Any additional intermediate-range correlations are too weak and/or too few to show. If there are additional \ch{Sn2S6^4-} ions associated to the NC beyond the bound layer, 
the ddPDF is insensitive to them. This finding is supported by the fact that the ddPDF does not change when additional multivalent salt is added to the solution (\textbf{Figure \ref{fig:variations}b}). We speculate that further layers of ions are 
present based on previous simulations,\cite{guerrero_garcia_polarization_2014,coropceanu_self-assembly_2022,jeong_atomic_2026} but are too disordered to be resolved in the ddPDF.

As shown in \textbf{Figure \ref{fig:modelvars}}, the calculated differential PDF is dominated by Pb-Sn correlations. While it is not surprising that the ddPDF would be dominated by the correlations between the strongest scattering atom type in the NC and the strongest scattering atom type in the ligands, it highlights the importance of the nanocrystal lattice structure in determining the ddPDF features. ddPDF studies have largely focused on the importance of the size and shape of the molecules in NC surface environments.\cite{zobel_universal_2015, kamysbayev_nanocrystals_2019}
We find in the electrostatically-stabilized PbS NC system that periodicity of the lattice, i.e., the positions of the Pb atoms contributing to the Pb-Sn pairs, plays at least as significant a role in setting the positions of the ddPDF features as the size and structure of the ligand molecules. Compared to typical systems of NCs with organic ligands, these strongly scattering atoms position us well to reveal an additional contribution to 
the periodicity seen in ddPDFs of colloidal NCs.

We support the idea that NC lattice periodicity contributes to the $\sim$3 \AA-period oscillations in the ddPDF with two distinct findings: the ddPDF measured for lower-Z 
CdSe NCs with \ch{Sn2S6^4-} ligands and the calculated differential PDF of the PbS NC model with a single Pt atom adjacent to it in place of ligands. The measured ddPDF of the colloidal CdSe NCs is shown in \textbf{Figure \ref{fig:variations}c}. This ddPDF lacks the oscillatory features in the PbS NC ddPDF. These features may be absent because of either or both of the following explanations. Cd is a weaker scattering element than both Sn and Pb. NC surface correlations may thus not be resolvable, supporting the notion that the strongly scattering, periodic Pb atoms in the PbS NC are responsible for the ddPDF oscillations. It is also possible that the CdSe NC ligands may lack the well-defined configurations that the PbS NC ligands seem to have, owing to the lower dielectric constant of CdSe and the consequential suppression of image charges and ligand grafting density. This deficiency likely contributes to the inability of CdSe NCs with \ch{Sn2S6^4-} ligands to assemble into ordered superlattices.\cite{coropceanu_self-assembly_2022} 

\begin{figure*}[!t]
    \centering
    \includegraphics[width=18cm]{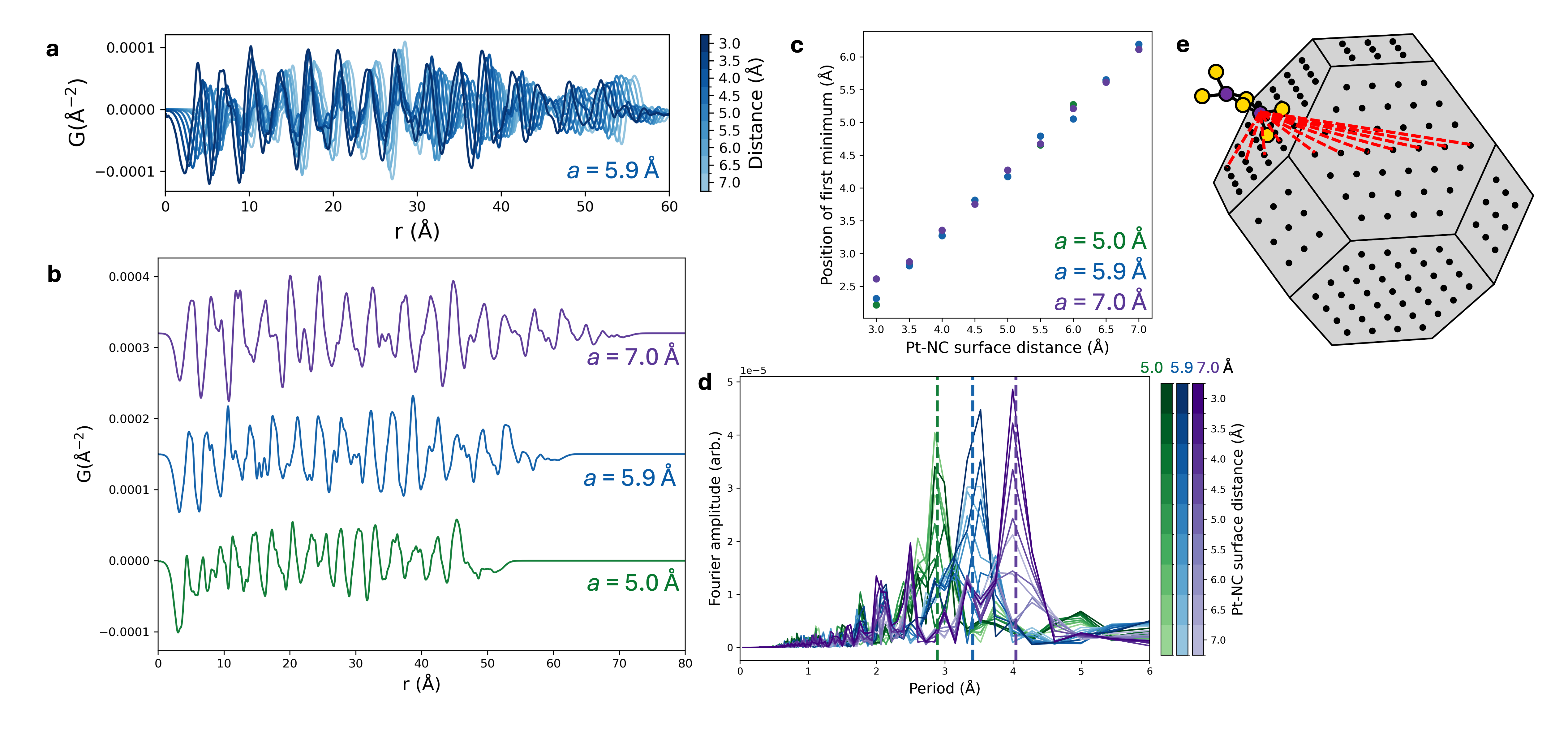}
    \caption[Oscillating differential PDFs of single Pt atom correlations with PbS NC.]{Oscillating differential PDFs of single Pt atom correlations with PbS NC.  \textbf{a.} Calculated differential PDFs with 0.7 $\mathrm{\AA}$ width Gaussian smearing for a single Pt atom near the surface of a PbS NC with the typical PbS lattice constant, $a$, of 5.9 $\mathrm{\AA}$. The distance of the Pt atom to the NC surface is varied from 3.0-7.0 $\mathrm{\AA}$. \textbf{b.} The 5.9 $\mathrm{\AA}$ lattice constant of the PbS NC is artificially increased and decreased to 7.0 $\mathrm{\AA}$ and 5.0 $\mathrm{\AA}$, respectively. The differential PDFs for a Pt-NC distance of 4.0 $\mathrm{\AA}$ are shown.  \textbf{c.} shows the onset point ($r$ value of first minimum) of the oscillations as a function of the Pt-NC surface distance for lattice constants, $a$, of 5.0, 5.9 (true PbS lattice constant), and 7.0 $\mathrm{\AA}$. All three lattice constants show the same linear trend. \textbf{d.} Fast Fourier transform spectra as a function of the PDF oscillation period for all three lattice constants explored and all Pt-NC surface distances. Dashed lines show $a/\sqrt{3}$ for each of the three lattice constants explored. The spectra show that the primary peak in the period increases with increased lattice constant but stays largely consistent with Pt-NC surface distance. \textbf{e} Illustration of the role of the NC lattice in ddPDFs. The correlations of the strongly scattering ligand atoms with the strongly scattering atoms in the lattice, both in the plane of the NC surface and throughout the depth of the NC, generate the oscillations in the ddPDF. 
}
    \label{fig:Ptmodel}
\end{figure*}

As a second element to support our hypothesis, we calculated a differential PDF for a model consisting of a single Pt atom near the surface of the same, but bare, PbS NC. This is a model system for testing our hypothesis about the origin of oscillations in the ddPDF and is not meant to be an accurate chemical model of our system. 
We choose Pt as a strongly scattering point particle with no internal structure. 
\textbf{Figure \ref{fig:Ptmodel}a} shows the differential PDFs (including partial PDFs of Pt-containing pairs, i.e., the proxy in our model of ligand-NC correlations, only) for distances between the Pt and the NC surface ranging between 3-7 $\mathrm{\AA}$. These PDFs show oscillations out to roughly the length scale of the NC diameter (5.8~nm). We repeated this calculation for NCs with artificially increased and decreased lattice constants (7.0~$\mathrm{\AA}$ and 5.0 $\mathrm{\AA}$, respectively). \textbf{Figure \ref{fig:Ptmodel}b} shows the differential PDFs for lattice constants of 5.0, 5.9 (true lattice constant of PbS), and 7.0 $\mathrm{\AA}$ and for a Pt-NC surface distance of 4.0 $\mathrm{\AA}$. Varying the distance of the atom from the NC surface linearly varies the position of the first minimum in the PDF, and it is independent of NC lattice constant, as shown in \textbf{Figure \ref{fig:Ptmodel}c}. The Fourier transforms of these PDFs
\cite{virtanen_scipy_2020} as a function of period in \textbf{Figure \ref{fig:Ptmodel}d} show that  
the dominant period of the signals increases with lattice constant and is largely independent of Pt-NC surface distance. Dashed lines show $a/\sqrt{3}$, the distance between Pb (111) planes, for each of the three lattice constants explored. This distance aligns with the peaks in the Fourier spectra. We thus demonstrate that a single platinum atom, which does not have a layered `solvation' structure or any internal molecular structure, exhibits oscillations in a differential PDF akin to those observed for colloidal NCs with a ligand/solvation shell. These oscillations extend beyond the distance between the atom and NC surface and decay on the length scale of the NC diameter. The onset point of these oscillations depends on the position of the atom but the period depends predominantly on the lattice constant. 

The simple Pt atom model highlights the key role that the NC lattice plays in determining the parameters of oscillations observed in the PbS NC ddPDF. Like the Pt atom, the \ch{Sn2S6^4-} ligands in the NC + ligand model and in the measured samples have correlations with the periodic Pb atoms in the NC that generate oscillations in the ddPDF. The interplanar spacing between Pb (111) layers is 3.4 $\mathrm{\AA}$, similar to the $\sim3\mathrm{\AA}$ period observed in \textbf{Figure \ref{fig:differences}c}. This finding identifies an additional contribution to consider when interpreting ddPDFs of colloidal NCs. Its relative importance likely depends on the atomic numbers of atoms in the NC and its surroundings, on how tightly the ligands are registered to their NC facets, and on the degree of order within the solvation shell. For example, in NC suspensions with ionic solvents that have stronger ion correlations, ion layering would likely dominate the ddPDF signal.\cite{kamysbayev_covalent_2020} 
The dominant role of the lattice also affects the sensitivity of a ddPDF to unbound species in the solvation shell. Because the presence of any successive layers of ions will be most strongly encoded in their correlations with the periodic Pb sublattice, they should contribute oscillations of similar period that differ mainly in onset. It follows that a ddPDF localizes a well-registered bound layer more readily than it resolves the number or arrangement of layers beyond it. We further generalize by calculating the differential PDF of an atom placed near a sphere with identical atoms on its unit grid points, showing that the oscillations also arise in this simple case (\textbf{Figure \ref{fig:grid}}).

We used ddPDF to measure the configuration of multivalent ligands on NCs stabilized electrostatically in multivalent electrolyte. Our ddPDFs show oscillations with a few-$\mathrm{\AA}$ period similar to the (111) Pb interplanar spacing in the NC lattice and that persist to $\sim$10 $\mathrm{\AA}$. 
We modeled this system with a truncated octahedral PbS NC, decorated with \ch{Sn2S6^4-} ligands. We identified the ligand radial positions, orientations, and their approximate distributions in calculated PDFs for the model ligand arrangement that best matches the experimental ddPDFs. From this model we inferred that ligands are arranged roughly perpendicular to the NC surface. Adding an additional layer of ions did not appreciably change the ddPDF on the length scale on which the PDF exhibits oscillations. 
These findings indicate that the ddPDF is dominated by the radially aligned, surface-bound ligand layer, and that it is only weakly sensitive to the arrangement of ions beyond it. 
From calculated partial PDFs we identified that the strong Pb-Sn correlations dominate the signal, highlighting the importance of the NC lattice in setting the features of the ddPDF. Furthermore, ddPDFs of more weakly scattering CdSe NCs with \ch{Sn2S6^4-} ligands exhibit minimal structure, supporting the notion that the strongly scattering, periodic Pb atoms in the PbS NC are responsible for the ddPDF oscillations. To further explore the role of the NC lattice in ddPDFs, we calculated PDFs of a simple model of a single Pt atom at a range of distances from the surface of PbS NCs with varied lattice constants. We showed that while the radial position of the onset of the observed oscillations depends on the distance of the Pt atom from the NC, the period of these oscillations depends on the lattice constant of the NC, supporting our hypothesis that the NC lattice plays a key role in determining the oscillation features in the PbS-\ch{Sn2S6^4-} ddPDF. Ultimately, our ddPDFs suggest radial alignment of \ch{Sn2S6^4-} ligands on electrostatically-stabilized PbS NCs, and we provide a new contribution to interpreting the ddPDF analysis of colloidally suspended NCs that emphasizes not only the arrangement of the molecules around NCs, but also the NC lattice structure.

In conjunction with other experimental and simulation efforts,\cite{tanner_enhancing_2025,tanner_origins_2025,jeong_atomic_2026,jeong_colloidal_2024} our findings allow us to move toward specifying the interactions between NCs and ultimately toward developing optimal self-assembly protocols for strongly electronically-coupled optoelectronic materials. Furthermore, our insight into the significant role that the NC lattice can play in generating oscillations in ddPDFs will facilitate the understanding of the molecular structure around nanoscale materials in a range of fields including catalysis,\cite{uchida_modular_2018} energy storage,\cite{guo_self-assembly_2016} biotechnology\cite{shin_covid-19_2020} and more.

\section*{Conflicts of Interest}
There are no conflicts to declare.

\section*{Acknowledgments}
This work was primarily supported by the Office of Basic Energy Sciences (BES), US Department of Energy (DOE) (award DE-SC0019375) and was also supported by the U.S. Department of Energy, Office of Science, Office of Workforce Development for Teachers and Scientists, Office of Science Graduate Student Research (SCGSR) program. The SCGSR program is administered by the Oak Ridge Institute for Science and Education for the DOE under contract number DE‐SC0014664. The NC synthesis and structural characterization was supported by the U.S. Department of Energy, Office of Science, Basic Energy Sciences, Materials Sciences and Engineering Division, under grant DE-SC0025256.  The primary PDF experiments were performed on APS beam time award (https://doi.org/10.46936/APS-196713/60018956) from the Advanced Photon Source, a U.S. Department of Energy (DOE) Office of Science user facility operated for the DOE Office of Science by Argonne National Laboratory under Contract No. DE-AC02-06CH11357. Supplementary PDF experiments used beamline 28-ID-1 of the National Synchrotron Light Source II, a U.S. Department of Energy (DOE) Office of Science User Facility operated for the DOE Office of Science by Brookhaven National Laboratory under Contract No. DE-SC0012704. S.J.L.B. acknowledges the support by the W. M. Keck Foundation under Award No. 1567502. V.R.K.W., A.Y., and C.P.N.T. were supported by the NSF Graduate Research Fellowship Program under Grant No. 2139433. D.T.L. was supported by an Alfred P. Sloan Research Fellowship. N.S.G. was supported by a David and Lucile Packard Foundation Fellowship for Science.

\begin{suppinfo}
Supporting Information: Methods including X-rays scattering measurements, data calibration and processing, PDF extraction, calculation of double difference PDFs, and model development; supporting figures S1-S8 showing supplemental NC sample characterization, additional PDF measurements, and model development details; supporting table S1 showing double difference morph parameters.

\end{suppinfo}
\bibliography{references}

\clearpage
\section*{For table of contents only}
\begin{figure}[H]
    \centering
    \includegraphics[width=8.25cm]{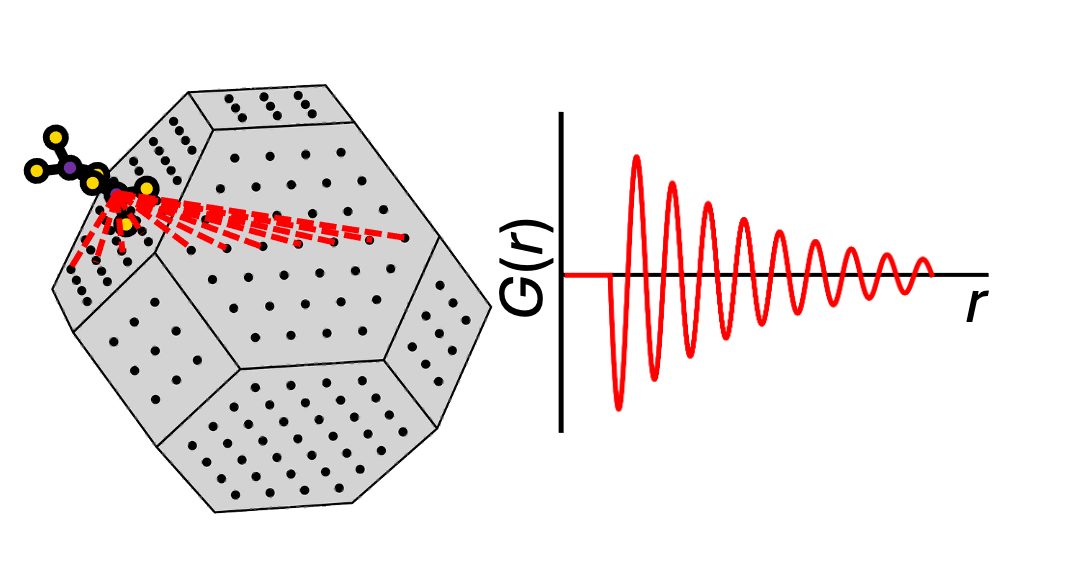}
    \label{fig:TOC}
\end{figure}


\end{document}